\documentclass[paper=a4, fontsize=11pt]{scrartcl}
\usepackage[T1]{fontenc}
\usepackage{fourier}

\usepackage[english]{babel}								
\usepackage[protrusion=true,expansion=true]{microtype}	
\usepackage{amsmath,amsfonts,amsthm}
\usepackage[pdftex]{graphicx}	
\usepackage{url}
\usepackage{siunitx}

\usepackage[pdfencoding=auto]{hyperref}
\usepackage{subcaption}
\usepackage[sorting=none,maxnames=6]{biblatex}
\usepackage{sectsty}
\allsectionsfont{ \normalfont} 

\usepackage{fancyhdr}
\numberwithin{equation}{section}		
\numberwithin{figure}{section}			
\numberwithin{table}{section}				

\newcommand{\horrule}[1]{\rule{\linewidth}{#1}} 	

\title{%
    \texorpdfstring{
		\usefont{OT1}{bch}{b}{n}
		\normalfont \normalsize \textsc{GSI Accelerator Physics Department} \\ [25pt]
		\horrule{0.5pt} \\[0.4cm]
		\huge Measurement of the Orbit Response Matrix for the SIS18 Doublet Optics \\
		\horrule{2pt} \\[0.5cm]
	}{%
	   Measurement of the Orbit Response Matrix for the GSI SIS18%
	}%
}
\newcommand{\aujordhui}{25. June 2024}
\author{%
    \texorpdfstring{
		\normalfont 								\normalsize
        Victoria Isensee, Conrad Caliari, 
        \normalsize\textit{TU Darmstadt, Germany} \\[-3pt]
        \normalsize
        Adrian Oeftiger, 
        \normalsize\textit{ GSI Helmholtz Centre for Heavy Ion Research GmbH, Darmstadt, Germany} \\[-3pt]
        \normalsize
        \aujordhui \\[1em]
        \normalsize
        GSI Report: GSI-2024-00625
    }{%
        Victoria Isensee, Conrad Caliari, Adrian Oeftiger%
    }%
}
\date{\texorpdfstring{}{\aujordhui}}

\begin{document}
\maketitle

\begin{abstract}
    \emph{Abstract:} This report describes the setup and results of a 2024 experiment on beam-based measurement of the Orbit Response Matrix (ORM) for the GSI Heavy Ion Synchrotron SIS18, characterising the standard doublet optics at extraction energy. The measured ORM is explicitly presented for reference.
    As an important outcome of this study, an effective linear machine model based on a MAD-X lattice could be established by employing the LOCO algorithm (Linear Optics from Closed Orbit), which fits free parameters in the model to match the measured ORM. The discussion compares the findings to previous measurements.
\end{abstract}

\section{Accelerator and Beam Properties}
\begin{figure}
    \centering
    \includegraphics[width=0.5\linewidth]{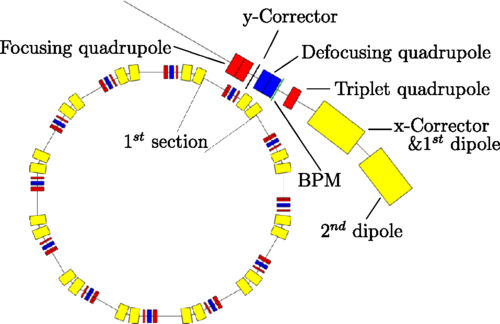}
    \caption{The schematic lattice of the SIS18 (Heavy Ion Synchrotron) shows the twelve sections of the ring accelerator. The first section is magnified and presents the labeled components. Source:\cite{CO_correction_circulant}}
    \label{fig:sis18_lattice}
\end{figure}

The 216-meter-long synchrotron SIS18 at GSI features twelve sections and is equipped with Beam Position Monitors (BPM) and steering magnets, which are used for closed orbit correction. The schematic layout of the SIS18 accelerator is depicted in Figure \ref{fig:sis18_lattice}\cite{CO_correction_circulant}. The first section, which has been magnified for more detailed observation in the top right corner of the figure, shows the labeled components: One section hosts one defocusing quadrupole surrounded by two focusing quadrupoles, where the latter is called triplet quadrupole.  If the strengths of the quadrupole elements are approximately equal, the configuration is referred to as a ``triplet optics''. In the alternative configuration, known as a ``doublet optics'', the triplet quadrupole has a strength that is approximately 10\% of that of the other two quadrupole elements. The defocusing and first focusing quadrupoles in odd-numbered sectors are designated as the odd family, while those in even-numbered sectors are designated as the even family. These families can be adjusted differently, i.e. to manipulate the transition energy. 
In the vertical plane, one BPM and one steerer are located in each section, situated at the same location. The horizontal steerers are located in the first dipole of all sections, with the exception of the fourth and sixth sections, where they are situated in the second dipole. 
The measurements were conducted in May 2024 using an $\mathrm{Ar}^{18+}$ beam. All measurements were performed at extraction energy and utilizing doublet optics. Using an intermediate beam process (after ``flat top''), a waiting period was maintained after the ramp until a quasi-equilibrium in terms of centroid oscillations was reached. Chromaticity was not corrected and left at the natural value, i.e., the sextupoles were deactivated, suppressing the associated nonlinear orbit response. A summary of the key parameters is provided in Table \ref{tab:beam_parameters}. 

\begin{table}[!h]
    \caption{Beam Parameters}
    \label{tab:beam_parameters}
\begin{center}
\begin{tabular}{ c c}
\hline
\hline
 Parameter & Value \\ 
\hline
 Ion & $\mathrm{Ar}^{18+}$ \\  
 Magnetic rigidity & 5.9 Tm  \\
 Betatron tunes $Q_x, Q_y$ & 4.29, 3.29 \\
 Bucket fill (Flat top) & 3 \\
\hline
\hline
\end{tabular}
\end{center}
\end{table}

\section{Description of the Measurement Process}

As a linear approximation, the Orbit Response Matrix (ORM) $\Omega$ 
\begin{equation}
    \Delta \vec{u}_{\mathrm{BPM}} = \Omega \Delta \vec{\theta}_{\mathrm{corr}}
    \label{eq:orm}
\end{equation}
describes how variations of the correction strength $\Delta \vec{\theta}_{\mathrm{corr}}$ 
affect changes in the trajectory of the particle beam, $\Delta \vec{u}_{\mathrm{BPM}}$. For the SIS18, $\Omega$ has $24\times 24$ entries and carries the unit mm/mrad. The vector $\Delta \vec{u}_{\mathrm{BPM}}=(\Delta x_1, ... \Delta x_{12}, \Delta y_1, ..., \Delta y_{12})^T$ contains the horizontal and vertical deviations $\Delta x$ and $\Delta y$ measured at the monitors in each of the two transverse planes. The vector $\vec{\theta}_{\mathrm{corr}}=(\theta_{\mathrm{corr, x1}}, ... \theta_{\mathrm{corr, x12}},\theta_{\mathrm{corr, y1}}, ...\theta_{\mathrm{corr, y12}})^T$ first incorporates the correction strengths in the horizontal direction, subsequently followed by the correction strengths in the vertical direction.
\begin{figure}[!h]
    \centering
    \includegraphics[width=0.65\linewidth]{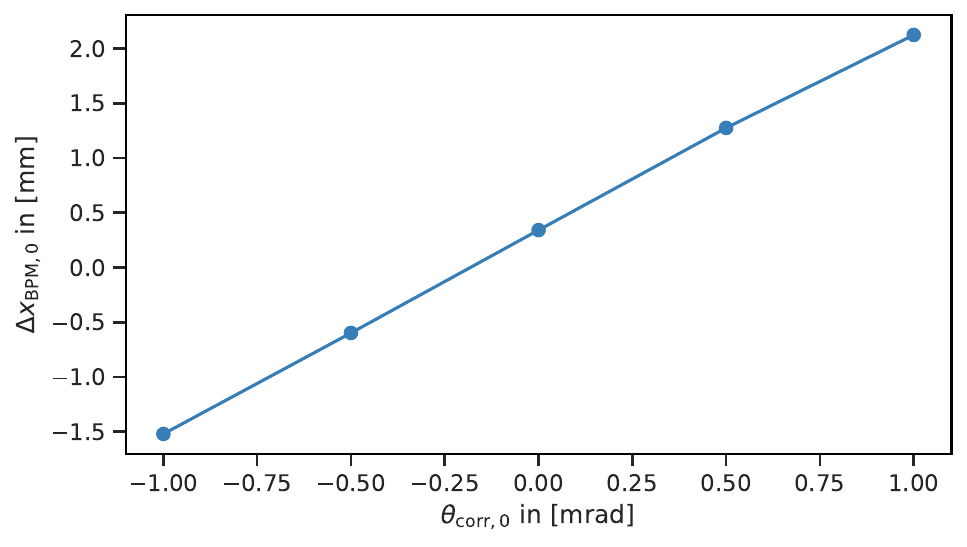}
    \caption{Measurement example: The influence of the primary horizontal steerer magnet (strength $\theta_{\mathrm{corr},x0}$) upon the first BPM (deviation $\Delta x_{\mathrm{BPM},0}$) in the horizontal plane. The resulting slope represents the corresponding individual ORM entry $\Omega_{i=0, j=0}$.}
    \label{fig:theta0_x0}
\end{figure}
The subsequent section will provide a detailed description of the measurement process. Each steering magnet, except for one, is initialized to a zero current, while a single steering magnet is powered at a predetermined non-zero value. The displacements are recorded at each BPM. Five consecutive shots are recorded due to fluctuations of the orbit. In order to reduce the influence of outliers and noise, the two measurements exhibiting the largest RMS deviation $|\Delta \vec{u}_{\mathrm{BPM}}|$ are excluded and the remaining three measurements are averaged. Then, the value of this particular steering magnet is changed, and the resulting displacements are recorded again. Subsequently, this procedure is repeated for the next steering magnet, where the previous magnet is reset to zero current while the subsequent magnet is assigned specified values. The specified values for the steerer strengths in this measurement are $\vec{\theta}_{\mathrm{corr}}= [-1, -0.5, 0, 0.5, 1]$\,mrad. Figure \ref{fig:theta0_x0} shows a measurement example of the strength $\theta_{x0}$ of the primary horizontal steering magnet and the horizontal deviation $\Delta x_0$ at the first BPM. After the described processing, the corresponding individual ORM entry $\Omega_{i=0, j=0}$ is the resulting slope.

\section{Measured Orbit Response Matrix}
With the described procedure the ORM was measured and its entries are displayed in Table \ref{tab:orm}.


\begin{table}[!h]
    \caption{The entries of the Orbit Response Matrix $\Omega$ from measurements and processing. The matrix is with $(\Delta \vec{x}_{\mathrm{BPM}}, \Delta \vec{y}_{\mathrm{BPM}})^T = \Omega \; (\vec{\theta}_{\mathrm{corr, x}}, \vec{\theta}_{\mathrm{corr, y}})^T$ comprised of 12 steerer magnets and 12 monitors in each of the two transverse planes, resulting in a size of
$24\times 24$ and has the unit mm/mrad. The subscripts are suppressed for notational simplicity.}
    \label{tab:orm}
\begin{center}
\resizebox{\textwidth}{!}{%
\begin{tabular} {|c||*{12}{c}|*{12}{c}|}
\hline 
 & $\theta_{x1}$& $\theta_{x2}$& $\theta_{x3}$& $\theta_{x4}$& $\theta_{x5}$& $\theta_{x6}$& $\theta_{x7}$& $\theta_{x8}$& $\theta_{x9}$& $\theta_{x10}$& $\theta_{x11}$& $\theta_{x12}$& $\theta_{y1}$& $\theta_{y2}$& $\theta_{y3}$& $\theta_{y4}$& $\theta_{y5}$& $\theta_{y6}$& $\theta_{y7}$& $\theta_{y8}$& $\theta_{y9}$& $\theta_{y10}$& $\theta_{y11}$& $\theta_{y12}$\\
\hline \hline
$\Delta x_1$ & 1.83 & 2.51 & -0.14 & -0.05 & 2.91 & -3.10 & -0.97 & 2.80 & -2.56 & 0.47 & 2.01 & -3.25 & -0.00 & 0.01 & 0.04 & 0.02 & 0.04 & 0.01 & 0.02 & 0.04 & 0.00 & -0.00 & -0.03 & 0.18 \\
$\Delta x_2$ & -3.02 & 1.84 & 2.53 & -2.57 & -2.24 & 2.56 & -1.66 & -0.95 & 2.93 & -2.68 & 0.50 & 2.25 & -0.01 & 0.06 & 0.00 & -0.03 & 0.01 & 0.06 & -0.01 & -0.03 & -0.04 & 0.06 & 0.02 & -0.01 \\
$\Delta x_3$ & 1.90 & -2.97 & 1.88 & 3.21 & -0.11 & 0.06 & 2.98 & -1.60 & -0.94 & 2.77 & -2.56 & 0.48 & 0.00 & -0.01 & -0.02 & 0.03 & 0.00 & -0.04 & 0.06 & 0.01 & -0.00 & -0.03 & -0.01 & 0.16 \\ 
$\Delta x_4$ & 0.60 & 1.91 & -3.02 & 3.32 & 2.33 & -2.49 & -2.04 & 2.93 & -1.62 & -0.81 & 2.72 & -2.70 & -0.01 & -0.02 & 0.04 & -0.01 & -0.06 & -0.02 & -0.02 & -0.00 & -0.01 & -0.02 & 0.01 & -0.14 \\
$\Delta x_5$ & -2.67 & 0.52 & 2.09 & -2.43 & 1.79 & 3.22 & -0.30 & -2.15 & 3.09 & -1.69 & -0.92 & 3.08 & 0.00 & 0.05 & -0.02 & -0.04 & 0.06 & 0.06 & -0.01 & -0.02 & -0.02 & 0.05 & 0.00 & -0.02 \\ 
$\Delta x_6$ & 2.88 & -2.58 & 0.39 & -0.16 & -2.90 & 3.20 & 2.48 & -0.13 & -2.33 & 3.05 & -1.51 & -1.23 & 0.03 & -0.04 & -0.03 & 0.08 & 0.04 & -0.02 & 0.02 & 0.01 & 0.04 & -0.03 & -0.04 & 0.15 \\
$\Delta x_7$ & -0.97 & 2.82 & -2.57 & 2.69 & 1.96 & -2.23 & 1.86 & 2.37 & -0.08 & -2.18 & 2.91 & -1.49 & -0.03 & 0.01 & 0.05 & -0.07 & -0.08 & 0.01 & -0.00 & -0.02 & -0.03 & 0.02 & 0.04 & -0.12 \\
$\Delta x_8$ & -1.61 & -1.01 & 2.94 & -3.26 & 0.43 & -0.35 & -3.11 & 1.94 & 2.46 & -0.20 & -2.18 & 3.26 & -0.01 & -0.04 & 0.00 & 0.02 & -0.04 & -0.03 & -0.04 & 0.02 & 0.01 & -0.05 & 0.02 & 0.00 \\
$\Delta x_{9}$ & 3.02 & -1.54 & -1.07 & 1.41 & -2.47 & 2.63 & 2.07 & -2.99 & 1.75 & 2.52 & -0.14 & -2.43 & -0.00 & 0.04 & -0.03 & -0.01 & 0.01 & 0.02 & 0.02 & -0.03 & -0.00 & 0.03 & 0.02 & -0.12 \\
$\Delta x_{10}$ & -2.20 & 3.03 & -1.55 & 1.51 & 2.76 & -3.00 & 0.52 & 1.93 & -2.96 & 1.82 & 2.43 & -0.09 & -0.00 & 0.00 & 0.04 & -0.02 & -0.01 & -0.01 & 0.00 & -0.01 & 0.04 & -0.01 & -0.01 & 0.09 \\
$\Delta x_{11}$ & -0.20 & -2.23 & 3.08 & -3.30 & -0.98 & 1.23 & -2.72 & 0.59 & 2.05 & -3.08 & 1.90 & 2.58 & -0.01 & 0.02 & -0.02 & -0.01 & 0.04 & 0.03 & -0.03 & 0.00 & -0.02 & 0.02 & 0.02 & 0.14 \\
$\Delta x_{12}$ & 2.45 & -0.14 & -2.27 & 2.65 & -1.45 & 1.51 & 2.89 & -2.60 & 0.44 & 2.06 & -3.02 & 1.95 & -0.02 & 0.00 & 0.01 & -0.01 & -0.06 & 0.01 & 0.06 & -0.06 & -0.01 & 0.03 & 0.00 & 0.05 \\ \hline
$\Delta y_1$ & -0.02 & -0.05 & -0.04 & -0.03 & -0.06 & 0.08 & 0.04 & -0.04 & -0.00 & -0.03 & -0.00 & 0.03 & 8.16 & 8.58 & -8.32 & -6.04 & 10.09 & 3.06 & -11.10 & 0.18 & 11.00 & -3.21 & -8.67 & 6.22 \\
$\Delta y_2$ & 0.03 & -0.09 & 0.03 & -0.00 & 0.03 & -0.09 & 0.01 & 0.05 & -0.09 & 0.13 & -0.05 & 0.03 & 6.31 & 7.95 & 8.62 & -8.11 & -6.18 & 9.84 & 3.30 & -10.89 & 0.15 & 10.84 & -2.93 & -9.61 \\
$\Delta y_3$ & 0.04 & 0.05 & -0.03 & 0.06 & 0.05 & -0.06 & -0.02 & 0.02 & 0.01 & 0.01 & 0.02 & -0.06 & -10.26 & 6.38 & 8.23 & 8.69 & -8.29 & -6.20 & 10.21 & 3.25 & -11.20 & -0.14 & 9.71 & -3.30 \\
$\Delta y_4$ & 0.03 & 0.01 & -0.02 & 0.09 & -0.01 & 0.08 & 0.06 & -0.06 & 0.04 & -0.07 & 0.01 & -0.07 & -3.23 & -9.93 & 6.31 & 7.99 & 8.73 & -8.01 & -6.40 & 9.97 & 3.25 & -10.85 & -0.01 & 10.73 \\
$\Delta y_5$ & -0.06 & -0.06 & 0.05 & -0.09 & -0.03 & 0.08 & 0.00 & -0.03 & 0.01 & -0.01 & -0.04 & 0.07 & 11.27 & -3.39 & -10.18 & 6.51 & 8.22 & 8.66 & -8.30 & -6.28 & 10.26 & 3.45 & -9.76 & 0.14 \\
$\Delta y_6$ & -0.03 & -0.00 & 0.04 & -0.11 & 0.04 & -0.02 & -0.06 & 0.03 & 0.00 & 0.05 & -0.02 & 0.09 & -0.05 & 10.81 & -3.33 & -9.81 & 6.23 & 7.91 & 8.75 & -8.06 & -6.18 & 9.76 & 2.81 & -10.51 \\
$\Delta y_7$ & 0.11 & -0.05 & 0.02 & 0.04 & -0.05 & -0.00 & 0.05 & -0.06 & -0.02 & 0.08 & -0.05 & -0.05 & -11.20 & 0.17 & 11.10 & -3.57 & -10.01 & 6.44 & 8.18 & 8.61 & -8.39 & -6.31 & 8.88 & 2.96 \\
$\Delta y_8$ & -0.05 & 0.02 & -0.00 & 0.04 & 0.01 & -0.01 & -0.02 & -0.01 & 0.06 & -0.10 & 0.02 & 0.00 & 3.28 & -10.79 & 0.08 & 10.76 & -3.28 & -9.73 & 6.17 & 7.95 & 8.55 & -7.85 & -5.34 & 9.59 \\
$\Delta y_9$ & -0.08 & 0.07 & -0.08 & 0.01 & 0.06 & -0.00 & 0.01 & 0.04 & -0.05 & -0.06 & 0.10 & -0.00 & 10.20 & 3.30 & -11.19 & 0.12 & 11.12 & -3.35 & -10.21 & 6.38 & 8.17 & 8.87 & -7.19 & -6.05 \\
$\Delta y_{10}$ & 0.01 & 0.01 & 0.03 & -0.05 & 0.03 & -0.05 & -0.02 & 0.06 & -0.09 & 0.07 & -0.02 & 0.02 & -6.11 & 9.96 & 2.98 & -10.88 & 0.28 & 10.72 & -3.43 & -9.76 & 6.51 & 7.96 & 7.36 & -7.97 \\
$\Delta y_{11}$ & -0.01 & -0.02 & 0.07 & 0.01 & 0.00 & -0.07 & -0.03 & 0.02 & 0.00 & 0.04 & -0.02 & 0.06 & -8.24 & -6.22 & 10.16 & 3.08 & -11.07 & 0.11 & 11.11 & -3.37 & -10.00 & 6.12 & 7.04 & 8.30 \\ 
$\Delta y_{12}$ & -0.01 & 0.00 & -0.07 & 0.07 & -0.03 & 0.05 & 0.05 & -0.06 & 0.06 & -0.06 & 0.02 & -0.02 & 8.80 & -8.05 & -6.26 & 9.96 & 3.24 & -10.82 & -0.09 & 10.91 & -3.38 & -9.98 & 5.52 & 7.71 \\
 \hline
\end{tabular}}
\end{center}
\end{table}

\begin{figure}[htbp]
   \begin{subfigure}{0.483\linewidth}
        \includegraphics[width=\linewidth]{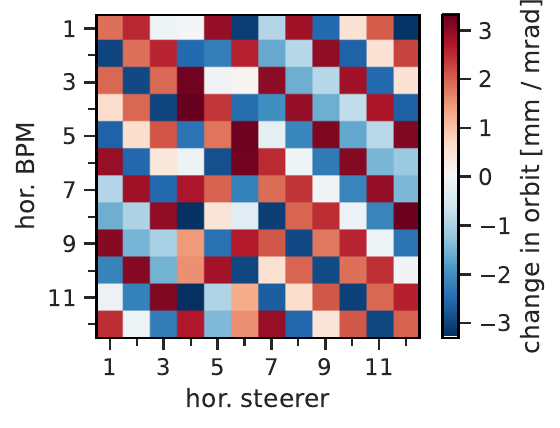}
        \caption{The measured horizontal submatrix.}
        \label{fig:submatrix_horizontal}
    \end{subfigure}
    \hfill
    \begin{subfigure}{0.496\linewidth}
        \includegraphics[width=\linewidth]{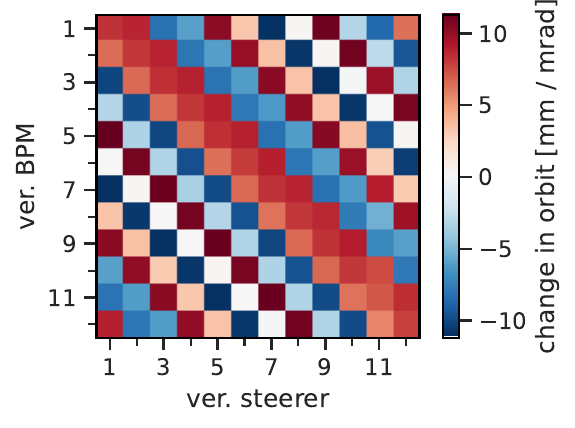}
        \caption{The measured vertical submatrix.}
        \label{fig:submatrix_vertical}
    \end{subfigure}
    \caption{Visualization of the submatrices of the ORM on the diagonal, representing horizontal steering vs.\ horizontal deviation, and vertical steering vs.\ vertical deviation.}
\end{figure}

The values displayed in the upper right quadrant of the ORM in Table \ref{tab:orm} show the relationship between the horizontal deviation at the BPMs and the strength of the vertical steerer magnets. In the lower left section, the relationship is vice versa, namely, the strength of the horizontal steerer magnets and the vertical deviation at the BPMs. These areas represent the extent of coupling between the transverse planes. As depicted in Figure \ref{fig:submatrix_vertical}, the lower right sub matrix illustrates that the vertical section of the ORM exhibits a circulant character. A circulant matrix is a matrix, where all row vectors are composed of the same elements and each row vector is rotated by one element to the right relative to the preceding row vector (for more information see Ref.~\cite{circulantMatrices}). Figure \ref{fig:submatrix_horizontal} visualizes the upper left submatrix of the ORM, which represents the horizontal dependencies. Apart from the fourth and sixth rows, where the displacement of the fourth and sixth steerer magnet becomes evident (as previously described in Section 1), the aforementioned circulant character is still evident. 

\section{LOCO \& Matching of the ORM}
A model ORM has been matched to the measured ORM by exploiting the LOCO algorithm (Linear Optics from Closed Orbits) \cite{safranek1997experimental}. A MAD-X model with the reference lattice of the GSI SIS18 synchrotron \cite{lattice} has been employed for the simulation. The matching parameters utilized in this process are the strengths of the focusing and defocusing quadrupole families for both the even and odd numbered sectors, as well as the triplet quadrupole magnets. These parameters were varied in the MAD-X model with the objective of minimizing the deviation between the model and the measured ORM. The initial integral quadrupole strengths $K_1\cdot L$ are summarized in Table \ref{tab:quad_parameter}. 
\begin{table}[h!]
\centering
\begin{tabular}{ l c } 
\hline
\hline
     & Initial Strengths \\
\hline
Focusing, odd    &  0.37 \\
Focusing, even   &  0.37 \\
Defocusing, odd  & -0.35 \\
Defocusing, even & -0.35 \\
Triplet          &  0.03 \\
\hline
\hline
\end{tabular}
 \caption[Quadrupole integral strengths for the even and odd quadrupole families]{Initial integral quadrupole strengths for the even and odd quadrupole families used for the simulation of the initial ORM and as initial values for the LOCO algorithm. The strengths correspond to the doublet optics configuration.}
 \label{tab:quad_parameter}
\end{table}
The Nelder-Mead method was chosen for the minimization process, with 5000 iterations as the maximum number to be performed. The algorithm converged prior to reaching the limit on the number of iterations.

\begin{figure}[h!]
    \centering
    \includegraphics[width=0.75\linewidth]{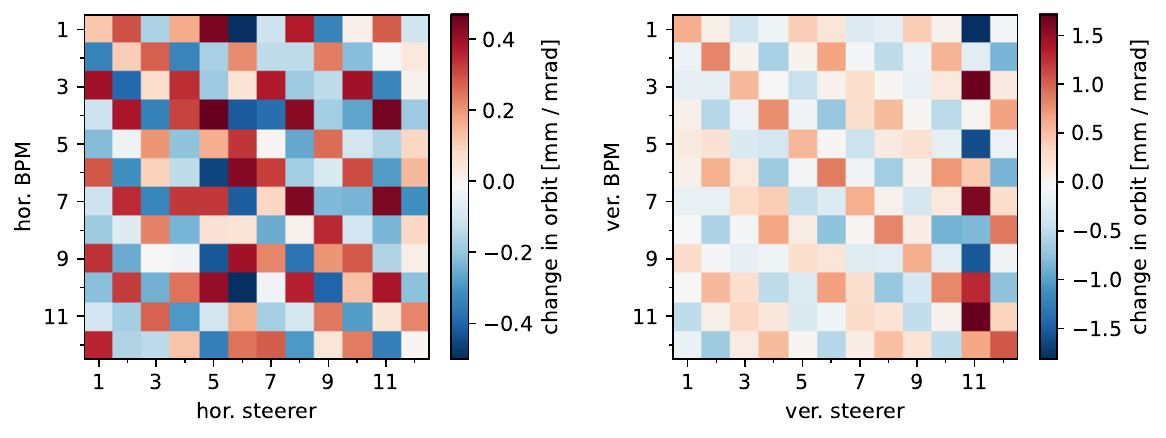}
    \caption{The discrepancy between the initial and measured ORM in the horizontal and vertical sections is illustrated.}
    \label{fig:init-exp}
\end{figure}
\begin{figure}[h!]
    \centering
    \includegraphics[width=0.8\linewidth]{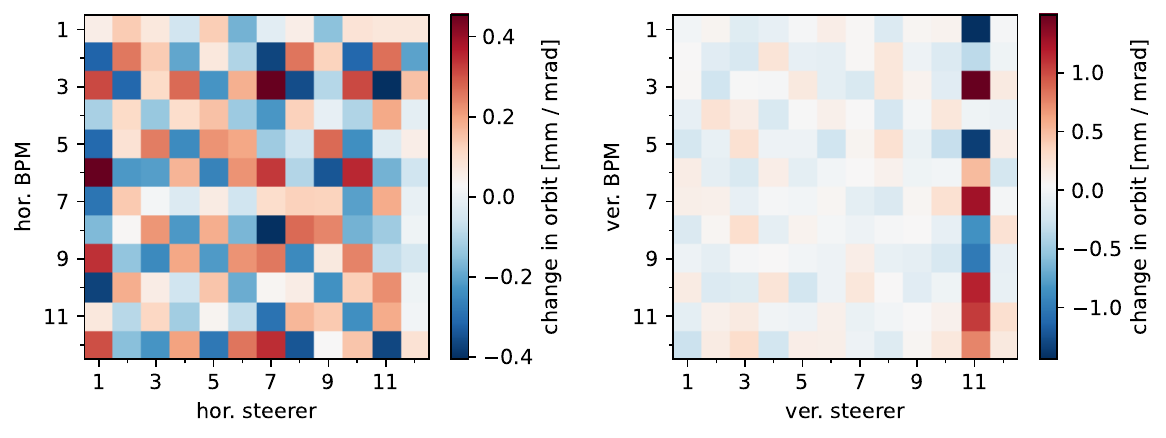}
    \caption{The difference between the matched and measured ORM in  the horizontal and vertical sections is visualized.}
    \label{fig:matched-exp}
\end{figure}
\begin{figure}[h!]
    \centering
    \includegraphics[width=0.8\linewidth]{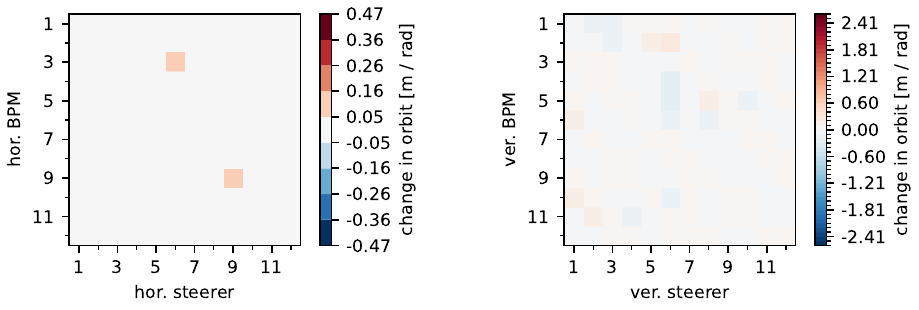}
    \caption{Residual discrepancy between measured and fitted ORM in case BPM roll angles and steerer gain errors are introduced as degrees of freedom.}
    \label{fig:extentedMatching}
\end{figure}

The matched integral quadrupole strengths, resulting from the LOCO process, are employed to simulate the matched ORM, which is then compared to the measured ORM. For purposes of comparison, the difference is calculated between the measured ORM and both the initial ORM and the matched ORM. Figure \ref{fig:init-exp} illustrates the discrepancy between the initial and measured ORM, while Figure \ref{fig:matched-exp} depicts the difference between the matched and measured ORM. As can be seen in both figures, the values for the vertical steerer in sector 11 (steerer S11KM2DV) are found to be considerably larger.

In order to further decrease differences between measured and simulated ORM BPM roll angles and deviations in steerer gain are introduced to the matching procedure, cf.\ Fig.~\ref{fig:extentedMatching}.
As a main result of LOCO, their predicted values are summarized in Table~\ref{Exp:Tab:SteererGainError_BPMRoll}.
Except for the BPM in Sector 8, no major BPM rotations along the reference trajectory are predicted.
The predicted gain errors indicate a systematic increase in horizontal and a systematic decrease in vertical steerer gain.
Moreover, the vertical steerer located in Sector 11, S11KM2DV, exhibits a large reduction in gain of 15\%.

\begin{table}[h]
    \centering
    \caption[BPM Roll Angle / Steerer Gain Error Predicted by LOCO]{BPM roll angle around the reference path and discrepancies in steerer gain predicted by LOCO.}
    \resizebox{\textwidth}{!}{%
        \begin{tabular}{c|c|c|c|c|c|c|c|c|c|c|c|c}
        \hline
        \hline
           no. sector  & 1 & 2 & 3 & 4 & 5 & 6 & 7 & 8 & 9 & 10 & 11 & 12 \\
           \hline
           \hline
           BPM roll [\si{\degree}]  & 0.02 & -0.1 & 0.09 & -0.08 & -0.03 & 0.02 & 0.88 & 6.24 & 0.19 & -0. & -0.07 & -0.17 \\
           \hline
           hor. gain [\si{\percent}] & 2.23 & 1.78 & 2.71 & 2.44 & 2.55 & 1.14 & 1.71 & 2.96 & 2.92 & 2.4 & 2.4 & 2.47 \\
           \hline
           ver. gain [\si{\percent}] & -1.78 & -1.74 & -3.43 & -2.66 & -3.98 & -5.53 & -2.55 & -2.78 & -2.71 & -2.99
    & -15.07 & -3. \\
           \hline
           \hline
        \end{tabular}
        }
    \label{Exp:Tab:SteererGainError_BPMRoll}
\end{table}

In her doctoral thesis, A.~Parfenova describes the measurements of three data sets collected in 2006 \cite{parfenova2008linear}. There, the three data sets were analyzed separately and fitted to a model on three separate occasions. The ORM modeling confirmed that all vertical steerers have a calibration factor of 1.0, but again with the exception of the steerer 11 (S11KM2DV), which exhibited a calibration factor of approximately 0.7. The ORM modeling yielded a calibration factor that was 33\% lower. It was determined that the magnetic field of this steerer is approximately 30\% weaker than anticipated. The thesis posited that the steerer S11KM2DV, in comparison to the other vertical steerers, has a modified design, namely a larger aperture and longer length. These modifications were necessary to provide a sufficiently large aperture in the area of re-injection. 
A subsequent hall probe measurement \cite{Muehle:VerticalSteererSIS18} indicated an affiliated reduction in field strength of 22\%. Although crosstalk effects with neighboring magnets were considered to be the likely due to the observed large reduction in steerer gain, a recalibration was performed based on the Hall probe measurement.

We conclude from our results that the reduction in steerer gain of S11KM2DV obtained by matching the measured ORM continues to provide a discrepancy between beam dynamical observations and Hall probe measurements, as of today (2024) there is still a miscalibration present.

\printbibliography

\end{document}